\documentclass[journal]{IEEEtran}

\usepackage[utf8]{inputenc}
\usepackage[T1]{fontenc} 

\usepackage{amsmath,amsfonts,amssymb}
\usepackage{algorithmic}
\usepackage{algorithm}
\usepackage{array}
\usepackage[caption=false,font=normalsize,labelfont=sf,textfont=sf]{subfig}
\usepackage{textcomp}
\usepackage{stfloats}
\usepackage{url}
\usepackage{verbatim}
\usepackage{graphicx}
\usepackage[colorlinks=true, urlcolor={blue!60!black}, linkcolor=black, citecolor=black]{hyperref}
\usepackage{csquotes}
\usepackage{enumitem}
\usepackage{tikz}
\usetikzlibrary{shapes.geometric, arrows, fit}

\usepackage{colortbl}
\usepackage{xcolor}
\usepackage{multirow}
\usepackage{booktabs}
\usepackage{tabularx}

\usepackage{cancel}
\usepackage[normalem]{ulem}

\renewcommand{\texttt}[1]{{\ttfamily\hyphenchar\font=`\-\relax #1}}

\usepackage[numbers,sort&compress]{natbib}

\newcommand*{\enq}[1]{\enquote{{\itshape#1}}}

\newcommand{\onlinetoolurl}{\url{https://se-uhd.de/ai-slop/}}
\newcounter{onlinetoolfn}

\newcommand{\supplementaryurl}{\url{https://doi.org/10.5281/zenodo.19283651}}
\newcounter{supplementaryfn}

\begin{document}

\title{``An Endless Stream of AI Slop'': How~Developers~Discuss~the Burden of AI-Assisted~Software~Development}

\author{
\IEEEauthorblockN{Sebastian Baltes\IEEEauthorrefmark{1}, Marc Cheong\IEEEauthorrefmark{2}, Christoph Treude\IEEEauthorrefmark{3}}\\
\IEEEauthorblockA{\IEEEauthorrefmark{1}Heidelberg University, Germany\\
\emph{sebastian.baltes@uni-heidelberg.de}}\\
\IEEEauthorblockA{\IEEEauthorrefmark{2}University of Melbourne, Australia\\
\emph{marc.cheong@unimelb.edu.au}}\\
\IEEEauthorblockA{\IEEEauthorrefmark{3}Singapore Management University, Singapore\\
\emph{ctreude@smu.edu.sg}}
}


\maketitle

\begin{abstract}
``AI slop'', that is, low-quality AI-generated content, is increasingly affecting software development, from generated code and pull requests to documentation and bug reports. However, there is limited empirical research on how developers perceive and respond to this phenomenon. We qualitatively analyzed how developers discuss AI slop in 1{,}154 Reddit and Hacker News posts, developing a codebook of 15 codes organized into three thematic clusters: \emph{Review Friction} (how AI slop burdens reviewers, erodes trust, and prompts countermeasures), \emph{Quality Degradation} (damage to codebases, knowledge resources, and developer competence), and \emph{Forces and Consequences} (systemic incentives, mandated adoption, craft erosion, and workforce disruption). Our findings frame AI slop as a \emph{tragedy of the commons}, where individual productivity gains externalize costs onto reviewers, maintainers, and the broader community. We report the concerns developers raise and the mitigation strategies they propose, with implications for tool developers, team leads, and educators.
\end{abstract}


\section{Introduction}
\label{sec:introduction}

``AI slop'' was named Merriam-Webster's 2025 Word of the Year.\footnote{\href{https://www.merriam-webster.com/slang/slop}{merriam-webster.com/slang/slop}} The term describes low-quality digital content produced, usually in quantity, with AI. Like ``spam'' before it, the term names a category of unwanted digital content. \citeauthor{kommers2025slop} identify three prototypical properties of AI slop~\cite{kommers2025slop}: \emph{superficial competence} (a veneer of quality belied by a deeper lack of substance), \emph{asymmetry of effort} (creation requires vastly less labor than would be needed without AI), and \emph{mass producibility} (content exists within a digital ecosystem of widespread generation and consumption). These properties separate AI slop from ordinary low-quality code and related terms: Slop is superficially competent, effort-asymmetric, and mass-producible. A hallucination, i.e., \enq{a plausible but false or misleading response generated by an artificial intelligence algorithm},\footnote{\href{https://www.merriam-webster.com/dictionary/hallucination}{merriam-webster.com/dictionary/hallucination}} names a correctness failure of model output---slop can be free of hallucinations and still waste its recipients' time. Spam, i.e., \enq{unsolicited usually commercial messages},\footnote{\href{https://www.merriam-webster.com/dictionary/spam}{merriam-webster.com/dictionary/spam}} serves a commercial or criminal interest. Slop requires no such motive and can arise from good-faith contributions such as pull requests or bug reports. Niederhoffer et al.\ coined the related term ``workslop'' for AI-generated content that destroys workplace productivity.\footnote{\href{https://hbr.org/2025/09/ai-generated-workslop-is-destroying-productivity}{hbr.org/2025/09/ai-generated-workslop-is-destroying-productivity}}

Much of the public discourse around AI slop has focused on social media\footnote{\href{https://www.9news.com.au/world/openai-sora-why-youre-seeing-ai-videos-on-your-social-media-feed-explained/d319cd50-a643-4c84-a542-a1913447508f}{9news.com.au/world/openai-sora-\ldots}} and search engine pollution.\footnote{\href{https://nymag.com/intelligencer/article/google-amazon-slop-internet.html}{nymag.com/intelligencer/\ldots/google-amazon-slop-internet}} However, AI slop has tangible impact on software development as well. It pervades code, bug reports, pull requests, documentation, and Stack Overflow answers. In open source, the problem is acute: \enq{AI slop is ripping up the social contract between maintainers and contributors essential to open source development}.\footnote{\href{https://redmonk.com/kholterhoff/2026/02/03/ai-slopageddon-and-the-oss-maintainers/}{redmonk.com/kholterhoff/2026/02/03/ai-slopageddon-\ldots}}

The consequences are already visible. The \emph{curl} project shut down its bug bounty program after AI-generated vulnerability reports consumed maintainer time without producing valid findings.\footnote{\href{https://lists.haxx.se/pipermail/daniel/2026-January/000143.html}{lists.haxx.se/pipermail/daniel/2026-January/000143.html}} Apache \emph{Log4j 2}\footnote{\href{https://github.com/apache/logging-log4j2/discussions/4052}{github.com/apache/logging-log4j2/discussions/4052}} and the Godot game engine\footnote{\href{https://bsky.app/profile/did:plc:v6vo5mafc2uirknvv5jegtzz/post/3meyerixvhs2p}{bsky.app/profile/\ldots/post/3meyerixvhs2p}} reported similar problems with AI-generated contributions that drained maintainer capacity. \citeauthor{koren2026vibecoding} argue more broadly that \emph{vibe coding} threatens the sustainability of open-source ecosystems by undermining maintainer incentives~\cite{koren2026vibecoding}.

This pattern resembles a \emph{tragedy of the commons}~\cite{hardin1968tragedy}, especially in open source where shared resources are maintained by volunteers. Individual developers and organizations benefit from AI-generated content, but the cumulative effect degrades the shared resources that collaborative development depends on. Codebases accumulate technical debt, knowledge resources become polluted, reviewer capacity is exhausted, and interpersonal trust erodes. Each AI-generated submission that skips quality review \emph{externalizes} its costs onto reviewers, maintainers, and the broader community. \citeauthor{prause2011reputation} previously applied this framing to documentation quality in software projects, arguing that documentation has low value to individual developers but high social cost when absent~\cite{prause2011reputation}.

Yet there is limited empirical research on how developers perceive and react to AI slop, even as practitioners report that mandated AI coding tools are \enq{rapidly becoming an endless stream of AI slop}~[\href{https://www.reddit.com/r/ExperiencedDevs/comments/1kr8clp/ai_slop_prs_are_burning_me_and_my_team_out_hard/}{R07}]. We study AI slop as a category in developer discourse, that is, how practitioners name, frame, and respond to it. We conducted a qualitative analysis of developer discourse on Reddit and Hacker News, guided by the following research question:

\begin{quote}
    \textbf{RQ:} \emph{How do software developers perceive and discuss ``AI slop''?}
\end{quote}

We analyzed 1{,}154 posts across 15 documents, developing a codebook of 15 codes organized into three thematic clusters: \emph{Review Friction}, \emph{Quality Degradation}, and \emph{Forces and Consequences}. Our findings capture the concerns developers raise and the strategies they propose, offering practical guidance for tool developers, team leads, and educators.

\section{Related Work}
\label{sec:relatedwork}

Prior work has examined AI-generated code quality, finding that LLM output frequently contains bugs and security vulnerabilities~\cite{pearce2022asleep, jesse2023large}. Surveys explore how developers perceive and adopt AI coding assistants as productivity aids~\cite{liang2024survey}.
Recent work analyzing Hacker News examined how AI-powered GitHub projects are promoted and received in developer communities~\cite{meakpaiboonwattana2025social}.
The degradation of online information quality through AI-generated content has been documented from model collapse on synthetic data~\cite{shumailov2024model} to incorrect AI-generated programming answers~\cite{kabir2024stackoverflow}. Research on open-source sustainability has long examined maintainer attention, maintenance burden, and burnout in collaborative development~\cite{eghbal2020working, raman2020stress}. AI slop introduces a new dimension by lowering the effort required to produce contributions while shifting the burden of quality assurance onto maintainers.

Our work differs from these studies in two ways. First, we study AI slop as a \emph{named} phenomenon, capturing how developers themselves frame the problem. Second, our qualitative approach captures the interplay between technical, social, and economic dimensions as they surface in practitioner discourse.

\section{Research Method}
\label{sec:researchmethod}

\subsection{Data Collection}

We collected discussions from \emph{Reddit}\footnote{\url{https://www.reddit.com}} and \emph{Hacker News}.\footnote{\url{https://news.ycombinator.com/}} On September 26--27, 2025, we searched for the exact phrase ``ai slop,'' with no date restriction, in three subreddits (\texttt{r/programming}, \texttt{r/learnprogramming}, \texttt{r/ExperiencedDevs}) and on Hacker News, where we narrowed the topical scope by adding the keyword ``software,'' because it has no equivalent of subreddits. For both sources, we screened all returned results and captured every discussion with at least one vote and one comment. This yielded 16 documents: 13 Reddit threads (\href{https://www.reddit.com/r/programming/comments/1n3brtd/ai_slop_attacks_on_the_curl_project_daniel/}{R01}, \href{https://www.reddit.com/r/programming/comments/1krh71o/github_wants_to_spam_open_source_projects_with_ai/}{R02}, \href{https://www.reddit.com/r/programming/comments/1hvr4gq/on_10_years_of_genai_slop_and_the_unfortunate/}{R03}, \href{https://www.reddit.com/r/learnprogramming/comments/1n12i6a/why_im_declining_your_ai_generated_mr/}{R04}, \href{https://www.reddit.com/r/ExperiencedDevs/comments/1nfpb14/reviewing_someone_elses_ai_slop/}{R05}, \href{https://www.reddit.com/r/ExperiencedDevs/comments/1mg2r6y/the_era_of_ai_slop_cleanup_has_begun/}{R06}, \href{https://www.reddit.com/r/ExperiencedDevs/comments/1kr8clp/ai_slop_prs_are_burning_me_and_my_team_out_hard/}{R07}, \href{https://www.reddit.com/r/ExperiencedDevs/comments/1m2yxdp/deal_with_ai_slop_at_c_level_execs/}{R08}, \href{https://www.reddit.com/r/ExperiencedDevs/comments/1l8ryy1/is_anyone_successfully_using_ai_assisted_coding/}{R09}, \href{https://www.reddit.com/r/ExperiencedDevs/comments/1l4n9jn/speaking_out_against_ai_fearmongering/}{R10}, \href{https://www.reddit.com/r/ExperiencedDevs/comments/1msywk0/are_programming_articlestutorials_and_docs/}{R11}, \href{https://www.reddit.com/r/ExperiencedDevs/comments/1ml0eme/folks_at_smmed_orgs_how_are_you_recruiting_these/}{R12}, \href{https://www.reddit.com/r/ExperiencedDevs/comments/1krttqo/my_new_hobby_watching_ai_slowly_drive_microsoft/}{R13}) and three Hacker News result pages (\href{https://hn.algolia.com/?dateRange=all\&page=0\&prefix=false\&query=\%22ai\%20slop\%22\%20software\&sort=byDate\&type=all}{H01}, \href{https://hn.algolia.com/?dateRange=all\&page=1\&prefix=false\&query=\%22ai\%20slop\%22\%20software\&sort=byDate\&type=all}{H02}, \href{https://hn.algolia.com/?dateRange=all\&page=2\&prefix=false\&query=\%22ai\%20slop\%22\%20software\&sort=byDate\&type=all}{H03}). One Reddit thread (\href{https://www.reddit.com/r/ExperiencedDevs/comments/1l8ryy1/is_anyone_successfully_using_ai_assisted_coding/}{R09}) was later excluded because it focused on successful AI use without engaging with AI slop as a problem. This left 15 documents (1{,}154 posts) in the final corpus.

We captured all threads and result pages as PDF files, which were the basis for our qualitative analysis. The earliest Reddit thread is from January 2025, and the cited Hacker News comments are from April to September 2025.

\subsection{Coding Procedure}

Our analysis followed an iterative qualitative coding approach. The second author performed open coding on five documents (R01--R05), generating 40 initial codes. All three authors then performed axial coding, grouping related open codes into 8 consolidated codes by comparing their meaning, scope, and overlap. The first author refined the codebook through ten revisions,\footnote{\label{fn:supplementary}\supplementaryurl}\setcounter{supplementaryfn}{\value{footnote}} assisted by Claude Code (Opus~4.6, high effort\footnote{We used the same tool/model to proofread and tighten this manuscript.}). Recent work cautions that claims about automating qualitative analysis with generative AI overgeneralize from narrow successes~\cite{ernst2026genai}, although studies have used LLM-assisted qualitative coding~\cite{dunivin2025scaling}. We report Claude Code's version, its role at each step, and the human verification of its output, following the community guidelines for empirical studies involving LLMs.\footnote{\href{https://llm-guidelines.org}{llm-guidelines.org}} The first author retained decision authority over all structural changes. The other two authors validated the final code set using an interactive visualization.\footnote{\label{fn:onlinetool}\onlinetoolurl}\setcounter{onlinetoolfn}{\value{footnote}}

The first author then annotated the full corpus with Claude Code in two passes (filtering for relevant sub-discussions, then coding) and reviewed the output at each step. Claude Code assigned the codes, but the authors held authority over every coding decision. Rather than checking only a sample of its labels, as is common, the second author then checked every label across all 15 documents over four review rounds,\footnotemark[\value{onlinetoolfn}] changing 234 of them. A detailed account is available in the online supplementary material.\footnotemark[\value{supplementaryfn}]$^,$\footnotemark[\value{onlinetoolfn}]

\subsection{Final Codebook}

\begin{figure}[t]
  \centering
  \includegraphics[width=\columnwidth]{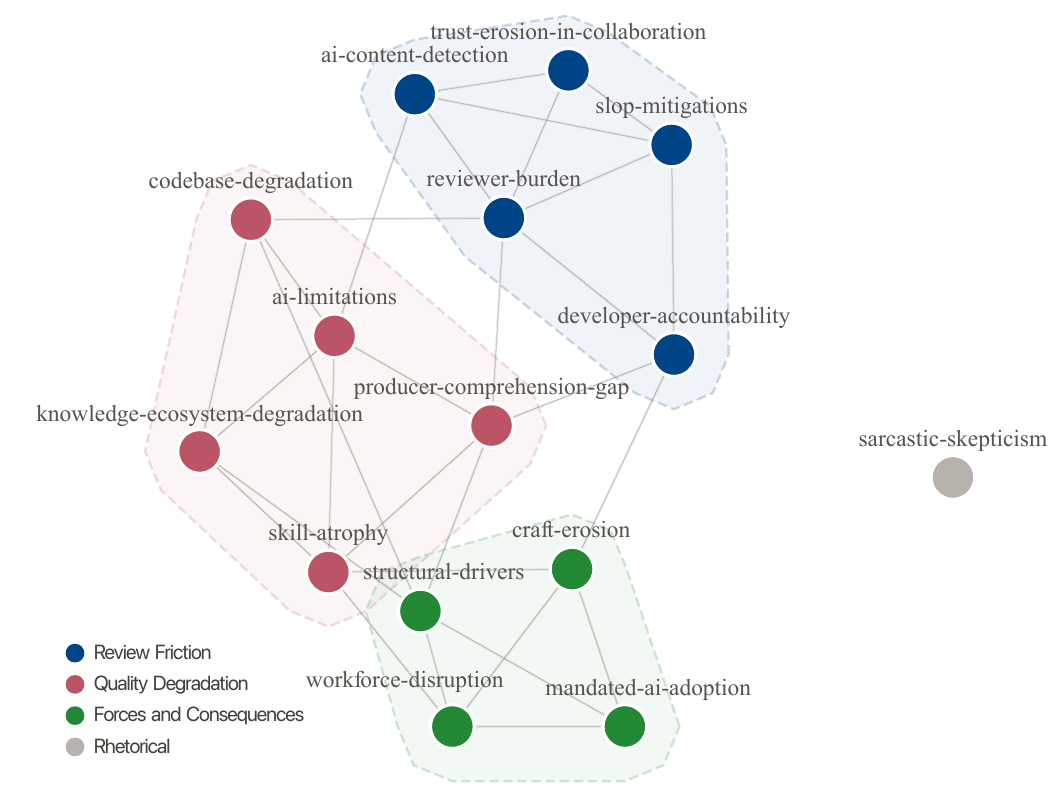}
  \caption{Code relationship network with Louvain community clusters. Edges represent conceptual relationships between codes, defined by the authors, capturing causal links, scope distinctions, and thematic overlaps. Node colors indicate cluster membership. An interactive version is available online.\protect\footnotemark[\value{onlinetoolfn}]}
  \label{fig:network}
\end{figure}

The final codebook consists of 15 codes: one \emph{rhetorical} code (\texttt{sarcastic-skepticism}) and 14 \emph{topical} codes. During the codebook development, we recorded relationships between codes as a graph whose nodes are codes and whose edges are conceptual links the authors identified by hand, capturing causal connections, scope distinctions, and thematic overlaps. We applied Louvain community detection, which partitions a graph by maximizing modularity, to identify three thematic clusters (\autoref{fig:network}), which we use to organize the presentation of results: \emph{Review Friction}, \emph{Quality Degradation}, and \emph{Forces and Consequences}. The codebook, corpus, and all annotated data are available online.\footnotemark[\value{supplementaryfn}]$^,$\footnotemark[\value{onlinetoolfn}]

\section{Results}
\label{sec:results}

We present findings along the three topical clusters (\autoref{tab:codebook}), plus the rhetorical code \texttt{sarcastic-skepticism}. The source IDs (e.g., \href{https://www.reddit.com/r/programming/comments/1krh71o/github_wants_to_spam_open_source_projects_with_ai/mtgx0fx/}{R02}) link to the original online posts.

\begin{table*}
  \caption{Codebook of 15 codes organized into three topical and one rhetorical clusters. Source IDs link to the posts on Reddit and Hacker News that motivated the corresponding code. The full codebook is available in an interactive online tool.\textsuperscript{\ref{fn:onlinetool}}}
  \label{tab:codebook}
  \centering
  \footnotesize
  \setlength{\tabcolsep}{3pt}
  \begin{tabularx}{\textwidth}{@{}>{\raggedright\arraybackslash}p{3.4cm}X>{\raggedright\arraybackslash}p{6.7cm}>{\raggedright\arraybackslash}p{1.9cm}@{}}
    \toprule
    \textbf{Code} & \textbf{Description} & \textbf{Representative Quote} & \textbf{Examples} \\
    \midrule
    \multicolumn{4}{@{}l}{\emph{Cluster:} \textbf{Review Friction} (5 codes)} \\
    \midrule
    {\ttfamily ai-\allowbreak content-\allowbreak detection} & Recognizing, flagging, or proving that content is AI-generated. & \enq{if the comment has an emoji it's a guarantee} & {\scriptsize \href{https://www.reddit.com/r/ExperiencedDevs/comments/1nfpb14/reviewing_someone_elses_ai_slop/}{R05} \href{https://www.reddit.com/r/ExperiencedDevs/comments/1nfpb14/reviewing_someone_elses_ai_slop/ne0qcdx/}{R05} \href{https://www.reddit.com/r/ExperiencedDevs/comments/1mg2r6y/the_era_of_ai_slop_cleanup_has_begun/}{R06} \href{https://www.reddit.com/r/ExperiencedDevs/comments/1mg2r6y/the_era_of_ai_slop_cleanup_has_begun/n6llkqf/}{R06}} \\
    {\ttfamily reviewer-\allowbreak burden} & Workload asymmetry where AI saves the author time while the reviewer bears the cost. & \enq{The development time has been shortened but the team now needs to spend more time to review. Doesn't look like any benefit.} & {\scriptsize \href{https://www.reddit.com/r/ExperiencedDevs/comments/1nfpb14/reviewing_someone_elses_ai_slop/ndy5dwv/}{R05} \href{https://www.reddit.com/r/ExperiencedDevs/comments/1nfpb14/reviewing_someone_elses_ai_slop/ndyb4bp/}{R05} \href{https://www.reddit.com/r/ExperiencedDevs/comments/1nfpb14/reviewing_someone_elses_ai_slop/ndyb97t/}{R05} \href{https://www.reddit.com/r/ExperiencedDevs/comments/1nfpb14/reviewing_someone_elses_ai_slop/ndyy932/}{R05} \href{https://www.reddit.com/r/ExperiencedDevs/comments/1nfpb14/reviewing_someone_elses_ai_slop/ne0hwyw/}{R05} \href{https://www.reddit.com/r/ExperiencedDevs/comments/1krttqo/my_new_hobby_watching_ai_slowly_drive_microsoft/mtg95rn/}{R13} \href{https://www.reddit.com/r/ExperiencedDevs/comments/1kr8clp/ai_slop_prs_are_burning_me_and_my_team_out_hard/}{R07}} \\
    {\ttfamily trust-\allowbreak erosion-\allowbreak in-\allowbreak collaboration} & Degraded trust in contributors or collaborative processes due to AI content. & \enq{When the comment smells like AI but I just can't prove it} & {\scriptsize \href{https://www.reddit.com/r/programming/comments/1krh71o/github_wants_to_spam_open_source_projects_with_ai/mtgx0fx/}{R02} \href{https://www.reddit.com/r/learnprogramming/comments/1n12i6a/why_im_declining_your_ai_generated_mr/nbkz0iv/}{R04} \href{https://www.reddit.com/r/ExperiencedDevs/comments/1krttqo/my_new_hobby_watching_ai_slowly_drive_microsoft/mtg727d/}{R13} \href{https://www.reddit.com/r/ExperiencedDevs/comments/1msywk0/are_programming_articlestutorials_and_docs/n9cbimf/}{R11} \href{https://www.reddit.com/r/ExperiencedDevs/comments/1ml0eme/folks_at_smmed_orgs_how_are_you_recruiting_these/n7n61jp/}{R12}} \\
    {\ttfamily developer-\allowbreak accountability} & Normative principle: developers bear full responsibility for submitted code, regardless of AI. & \enq{It's not AI's code, it's my code} & {\scriptsize \href{https://www.reddit.com/r/ExperiencedDevs/comments/1nfpb14/reviewing_someone_elses_ai_slop/ndyv6vr/}{R05} \href{https://www.reddit.com/r/ExperiencedDevs/comments/1nfpb14/reviewing_someone_elses_ai_slop/ndzdrgd/}{R05} \href{https://www.reddit.com/r/ExperiencedDevs/comments/1nfpb14/reviewing_someone_elses_ai_slop/ne4c1q4/}{R05} \href{https://www.reddit.com/r/ExperiencedDevs/comments/1nfpb14/reviewing_someone_elses_ai_slop/ndzkayp/}{R05}} \\
    {\ttfamily slop-\allowbreak mitigations} & Concrete actions taken or proposed to reduce, contain, or respond to AI slop. & \enq{less than 500 LOC per PR or they won't review it} & {\scriptsize \href{https://www.reddit.com/r/learnprogramming/comments/1n12i6a/why_im_declining_your_ai_generated_mr/nbkz0iv/}{R04} \href{https://www.reddit.com/r/ExperiencedDevs/comments/1nfpb14/reviewing_someone_elses_ai_slop/ndy44nt/}{R05} \href{https://www.reddit.com/r/ExperiencedDevs/comments/1nfpb14/reviewing_someone_elses_ai_slop/ndzk0xn/}{R05} \href{https://www.reddit.com/r/ExperiencedDevs/comments/1nfpb14/reviewing_someone_elses_ai_slop/ndzk6yl/}{R05} \href{https://www.reddit.com/r/ExperiencedDevs/comments/1nfpb14/reviewing_someone_elses_ai_slop/nebtt4d/}{R05}} \\
    \midrule
    \multicolumn{4}{@{}l}{\emph{Cluster:} \textbf{Quality Degradation} (5 codes)} \\
    \midrule
    {\ttfamily ai-\allowbreak limitations} & Poor output quality attributed to what the AI tool itself cannot do, independent of user skill. & \enq{almost all code ever written is mediocre (or worse) and that's what LLMs have been trained to replicate without understanding} & {\scriptsize \href{https://www.reddit.com/r/ExperiencedDevs/comments/1nfpb14/reviewing_someone_elses_ai_slop/ndzdefe/}{R05} \href{https://www.reddit.com/r/programming/comments/1hvr4gq/on_10_years_of_genai_slop_and_the_unfortunate/m5xbkl7/}{R03} \href{https://www.reddit.com/r/ExperiencedDevs/comments/1mg2r6y/the_era_of_ai_slop_cleanup_has_begun/n6m4wb6/}{R06} \href{https://www.reddit.com/r/ExperiencedDevs/comments/1mg2r6y/the_era_of_ai_slop_cleanup_has_begun/n6olboh/}{R06} \href{https://www.reddit.com/r/ExperiencedDevs/comments/1mg2r6y/the_era_of_ai_slop_cleanup_has_begun/n6pksg6/}{R06} \href{https://www.reddit.com/r/ExperiencedDevs/comments/1krttqo/my_new_hobby_watching_ai_slowly_drive_microsoft/mtgyqgi/}{R13} \href{https://www.reddit.com/r/ExperiencedDevs/comments/1krttqo/my_new_hobby_watching_ai_slowly_drive_microsoft/mtgyrxe/}{R13} \href{https://www.reddit.com/r/ExperiencedDevs/comments/1krttqo/my_new_hobby_watching_ai_slowly_drive_microsoft/mtgmow5/}{R13} \href{https://www.reddit.com/r/ExperiencedDevs/comments/1kr8clp/ai_slop_prs_are_burning_me_and_my_team_out_hard/}{R07}} \\
    {\ttfamily codebase-\allowbreak degradation} & Degraded technical quality within a project or codebase as a consequence of AI slop. & \enq{You can go very fast with AI, but you accrue technical debt at a much higher speed too} & {\scriptsize \href{https://www.reddit.com/r/ExperiencedDevs/comments/1mg2r6y/the_era_of_ai_slop_cleanup_has_begun/n6lnlpy/}{R06} \href{https://www.reddit.com/r/ExperiencedDevs/comments/1mg2r6y/the_era_of_ai_slop_cleanup_has_begun/n6lq3lb/}{R06} \href{https://www.reddit.com/r/ExperiencedDevs/comments/1nfpb14/reviewing_someone_elses_ai_slop/ndyb97t/}{R05} \href{https://www.reddit.com/r/learnprogramming/comments/1n12i6a/why_im_declining_your_ai_generated_mr/nbkz0iv/}{R04} \href{https://www.reddit.com/r/ExperiencedDevs/comments/1krttqo/my_new_hobby_watching_ai_slowly_drive_microsoft/mtgfqlp/}{R13} \href{https://www.reddit.com/r/ExperiencedDevs/comments/1kr8clp/ai_slop_prs_are_burning_me_and_my_team_out_hard/mtc7k6g/}{R07} \href{https://news.ycombinator.com/item?id=43624576}{H03}} \\
    {\ttfamily knowledge-\allowbreak ecosystem-\allowbreak degradation} & Degraded quality of external knowledge resources (docs, tutorials, Q\&A sites). & \enq{I'm starting to see documentation and tutorials missing key information and code samples needed to be able to implement something now} & {\scriptsize \href{https://www.reddit.com/r/ExperiencedDevs/comments/1msywk0/are_programming_articlestutorials_and_docs/}{R11} \href{https://www.reddit.com/r/ExperiencedDevs/comments/1msywk0/are_programming_articlestutorials_and_docs/n9bdsps/}{R11} \href{https://www.reddit.com/r/ExperiencedDevs/comments/1msywk0/are_programming_articlestutorials_and_docs/n98pdil/}{R11} \href{https://www.reddit.com/r/ExperiencedDevs/comments/1msywk0/are_programming_articlestutorials_and_docs/}{R11}} \\
    {\ttfamily producer-\allowbreak comprehension-\allowbreak gap} & Producers who lack the understanding needed to evaluate their own AI-generated output. & \enq{I straight up asked them if they know what their code does. They didn't. The PR was not approved.} & {\scriptsize \href{https://www.reddit.com/r/programming/comments/1krh71o/github_wants_to_spam_open_source_projects_with_ai/mti2e8p/}{R02} \href{https://www.reddit.com/r/programming/comments/1n3brtd/ai_slop_attacks_on_the_curl_project_daniel/nbk3cbr/}{R01} \href{https://www.reddit.com/r/ExperiencedDevs/comments/1nfpb14/reviewing_someone_elses_ai_slop/ndyi7tz/}{R05} \href{https://www.reddit.com/r/ExperiencedDevs/comments/1mg2r6y/the_era_of_ai_slop_cleanup_has_begun/n6lwgy1/}{R06}} \\
    {\ttfamily skill-\allowbreak atrophy} & Declining developer capability over time due to AI reliance. & \enq{many in our field are happily contributing to that future themselves by letting AI do their own job and slowly let their brains rot away} & {\scriptsize \href{https://www.reddit.com/r/programming/comments/1krh71o/github_wants_to_spam_open_source_projects_with_ai/mti2e8p/}{R02} \href{https://www.reddit.com/r/ExperiencedDevs/comments/1mg2r6y/the_era_of_ai_slop_cleanup_has_begun/n6m17tx/}{R06} \href{https://www.reddit.com/r/ExperiencedDevs/comments/1nfpb14/reviewing_someone_elses_ai_slop/ne0nx3v/}{R05} \href{https://www.reddit.com/r/ExperiencedDevs/comments/1l4n9jn/speaking_out_against_ai_fearmongering/mwd7evv/}{R10} \href{https://news.ycombinator.com/item?id=44049529}{H02}} \\
    \midrule
    \multicolumn{4}{@{}l}{\emph{Cluster:} \textbf{Forces and Consequences} (4 codes)} \\
    \midrule
    {\ttfamily structural-\allowbreak drivers} & Structural forces (incentives, corporate actors, cost-cutting) that drive AI slop production. & \enq{This will just be weaponised by people trying to boost their profile in an ironically shrinking job market.} & {\scriptsize \href{https://www.reddit.com/r/programming/comments/1krh71o/github_wants_to_spam_open_source_projects_with_ai/mtgx0fx/}{R02} \href{https://www.reddit.com/r/ExperiencedDevs/comments/1nfpb14/reviewing_someone_elses_ai_slop/ndzcpsz/}{R05} \href{https://www.reddit.com/r/ExperiencedDevs/comments/1nfpb14/reviewing_someone_elses_ai_slop/ndye2fl/}{R05} \href{https://www.reddit.com/r/ExperiencedDevs/comments/1kr8clp/ai_slop_prs_are_burning_me_and_my_team_out_hard/}{R07} \href{https://news.ycombinator.com/item?id=44732364}{H02} \href{https://www.reddit.com/r/programming/comments/1krh71o/github_wants_to_spam_open_source_projects_with_ai/mtez7pf/}{R02} \href{https://www.reddit.com/r/ExperiencedDevs/comments/1nfpb14/reviewing_someone_elses_ai_slop/ndydq6l/}{R05} \href{https://www.reddit.com/r/ExperiencedDevs/comments/1mg2r6y/the_era_of_ai_slop_cleanup_has_begun/n6mifky/}{R06} \href{https://news.ycombinator.com/item?id=44468293}{H02}} \\
    {\ttfamily mandated-\allowbreak ai-\allowbreak adoption} & AI features or workflows imposed on developers without meaningful choice. & \enq{[There has been] a huge push for teams to adopt tools like Cursor [\ldots] rapidly becoming an endless stream of AI slop.} & {\scriptsize \href{https://www.reddit.com/r/programming/comments/1krh71o/github_wants_to_spam_open_source_projects_with_ai/mtl1isq/}{R02} \href{https://www.reddit.com/r/ExperiencedDevs/comments/1mg2r6y/the_era_of_ai_slop_cleanup_has_begun/n6pvcvf/}{R06} \href{https://www.reddit.com/r/ExperiencedDevs/comments/1mg2r6y/the_era_of_ai_slop_cleanup_has_begun/n6u86mc/}{R06} \href{https://www.reddit.com/r/ExperiencedDevs/comments/1m2yxdp/deal_with_ai_slop_at_c_level_execs/}{R08} \href{https://www.reddit.com/r/ExperiencedDevs/comments/1kr8clp/ai_slop_prs_are_burning_me_and_my_team_out_hard/}{R07}} \\
    {\ttfamily craft-\allowbreak erosion} & Loss, grief, or disillusionment about what software development has become. & \enq{What does saving an hour or two writing code, which I actually like, give me if I have to spend that same hour or two deshittifying it?} & {\scriptsize \href{https://www.reddit.com/r/ExperiencedDevs/comments/1nfpb14/reviewing_someone_elses_ai_slop/ndyb4bp/}{R05} \href{https://www.reddit.com/r/ExperiencedDevs/comments/1nfpb14/reviewing_someone_elses_ai_slop/ndyuy0b/}{R05} \href{https://www.reddit.com/r/ExperiencedDevs/comments/1nfpb14/reviewing_someone_elses_ai_slop/ndyuilq/}{R05} \href{https://www.reddit.com/r/ExperiencedDevs/comments/1mg2r6y/the_era_of_ai_slop_cleanup_has_begun/n6lq3lb/}{R06} \href{https://news.ycombinator.com/item?id=45361392}{H01} \href{https://news.ycombinator.com/item?id=43579158}{H03}} \\
    {\ttfamily workforce-\allowbreak disruption} & How AI slop affects the software workforce: jobs, hiring, or career trajectories. & \enq{Fake LinkedIn profiles, fake GitHub, fake resumes. In a few cases even fake people} & {\scriptsize \href{https://www.reddit.com/r/ExperiencedDevs/comments/1mg2r6y/the_era_of_ai_slop_cleanup_has_begun/n6lj94k/}{R06} \href{https://www.reddit.com/r/ExperiencedDevs/comments/1mg2r6y/the_era_of_ai_slop_cleanup_has_begun/n6lql0z/}{R06} \href{https://www.reddit.com/r/ExperiencedDevs/comments/1mg2r6y/the_era_of_ai_slop_cleanup_has_begun/n6nrsbo/}{R06} \href{https://www.reddit.com/r/programming/comments/1krh71o/github_wants_to_spam_open_source_projects_with_ai/mtdssyj/}{R02} \href{https://www.reddit.com/r/ExperiencedDevs/comments/1mg2r6y/the_era_of_ai_slop_cleanup_has_begun/n6lj94k/}{R06} \href{https://www.reddit.com/r/ExperiencedDevs/comments/1ml0eme/folks_at_smmed_orgs_how_are_you_recruiting_these/}{R12} \href{https://www.reddit.com/r/ExperiencedDevs/comments/1ml0eme/folks_at_smmed_orgs_how_are_you_recruiting_these/n7mz1yp/}{R12} \href{https://www.reddit.com/r/ExperiencedDevs/comments/1ml0eme/folks_at_smmed_orgs_how_are_you_recruiting_these/n7no3xh/}{R12}} \\
    \midrule
    \multicolumn{4}{@{}l}{\emph{Cluster:} \textbf{Rhetorical} (1 code)} \\
    \midrule
    {\ttfamily sarcastic-\allowbreak skepticism} & Irony, mock enthusiasm, or absurdist framing. Applied alongside topical codes. & \enq{the final boss: outsourced AI slop} & {\scriptsize \href{https://www.reddit.com/r/programming/comments/1n3brtd/ai_slop_attacks_on_the_curl_project_daniel/nbcacb1/}{R01} \href{https://www.reddit.com/r/programming/comments/1krh71o/github_wants_to_spam_open_source_projects_with_ai/mtdl87e/}{R02} \href{https://www.reddit.com/r/ExperiencedDevs/comments/1nfpb14/reviewing_someone_elses_ai_slop/ndybc9v/}{R05} \href{https://www.reddit.com/r/ExperiencedDevs/comments/1mg2r6y/the_era_of_ai_slop_cleanup_has_begun/n6n2cfz/}{R06}} \\
    \bottomrule
  \end{tabularx}
\end{table*}

\subsection{Annotation Overview}

We annotated all 1{,}154 posts across 15 documents using the final codebook. Of all 1{,}154 posts, 978 (84.7\%) received at least one code, yielding 1{,}603 codings. On average, these posts received 1.6 codes, showing that developers often address multiple themes in one post.

\begin{figure}[t]
  \centering
  \includegraphics[width=\columnwidth]{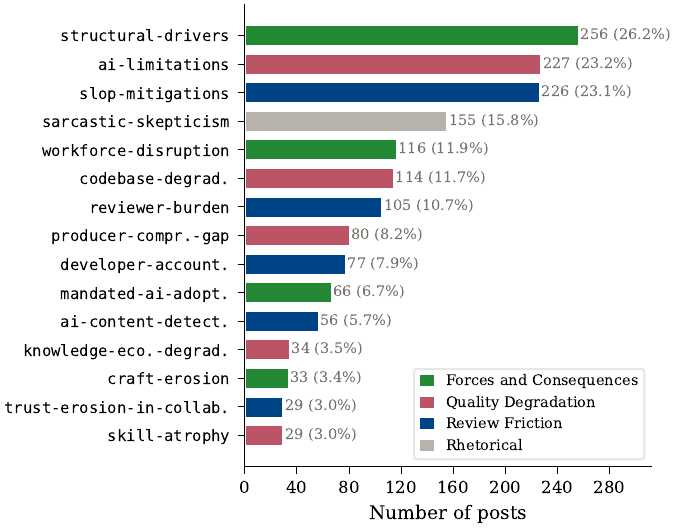}
  \caption{Code frequency distribution across 978 coded posts. Bar colors indicate cluster membership. An interactive version with co-occurrence data is available online.\protect\footnotemark[\value{onlinetoolfn}]}
  \label{fig:frequencies}
\end{figure}

\autoref{fig:frequencies} shows the frequency distribution. The three most frequent topical codes are \texttt{structural-drivers}~(256, or 26.2\% of coded posts), \texttt{ai-limitations}~(227), and \texttt{slop-mitigations}~(226), which together account for 44.2\% of all 1{,}603 codings. They reflect the fundamental questions: \emph{why} slop is produced, \emph{what} makes it problematic, and \emph{how} to counter it. The rhetorical code \texttt{sarcastic-skepticism} (155) ranks fourth, indicating that ironic framing was common in this discourse. Co-occurrence analysis shows cross-cutting patterns (\texttt{sarcastic-skepticism} pairs most often with \texttt{structural-drivers} and \texttt{ai-limitations}) and within-cluster pairings consistent with the Louvain clustering.

\subsection{Review Friction}

This cluster spans five codes addressing how AI slop disrupts collaborative development: detecting AI content (\texttt{ai-content-detection}), the workload it creates (\texttt{reviewer-burden}), the trust it erodes (\texttt{trust-erosion-in-collaboration}), and how developers respond through accountability norms (\texttt{developer-accountability}) and concrete countermeasures (\texttt{slop-mitigations}).

\textbf{Detection.}
Reviewers developed pattern recognition for AI-generated code. Explicit markers include emojis in code comments~[\href{https://www.reddit.com/r/ExperiencedDevs/comments/1mg2r6y/the_era_of_ai_slop_cleanup_has_begun/n6llkqf/}{R06}], step-by-step commenting, verbose style, and Unicode artifacts~[\href{https://www.reddit.com/r/ExperiencedDevs/comments/1nfpb14/reviewing_someone_elses_ai_slop/ne0qcdx/}{R05}]. Others relied on tacit recognition: \enq{Someone on my team will publish a PR, it's like 1--2k lines of code and after looking at it for 5 minutes I can tell it's pretty much entirely AI generated}~[\href{https://www.reddit.com/r/ExperiencedDevs/comments/1nfpb14/reviewing_someone_elses_ai_slop/}{R05}].

\textbf{Reviewer burden.}
The workload asymmetry between AI generation and human review was a dominant theme. \enq{The development time has been shortened but the team now needs to spend more time to review. Doesn't look like any benefit}~[\href{https://www.reddit.com/r/ExperiencedDevs/comments/1nfpb14/reviewing_someone_elses_ai_slop/ndy5dwv/}{R05}]. One team reported receiving 30 PRs per day across 6 reviewers~[\href{https://www.reddit.com/r/ExperiencedDevs/comments/1kr8clp/ai_slop_prs_are_burning_me_and_my_team_out_hard/}{R07}]. Reviewers described feeling like \enq{the first human being to ever lay eyes on this code}~[\href{https://www.reddit.com/r/ExperiencedDevs/comments/1nfpb14/reviewing_someone_elses_ai_slop/ndyb97t/}{R05}] and being turned into unpaid prompt engineers: \enq{They're literally just using you to do their job (i.e., critically evaluate and understand their AI slop and give it the next prompt)}~[\href{https://www.reddit.com/r/ExperiencedDevs/comments/1nfpb14/reviewing_someone_elses_ai_slop/ne0hwyw/}{R05}].

\textbf{Trust erosion.}
AI-generated contributions eroded trust. One reviewer described an AI agent's PR: \enq{I don't know how you could trust any of it [\ldots] No real understanding of what it's doing, it's just guessing}~[\href{https://www.reddit.com/r/ExperiencedDevs/comments/1krttqo/my_new_hobby_watching_ai_slowly_drive_microsoft/mtg727d/}{R13}]. Proving AI authorship is hard: \enq{When the comment smells like AI but I just can't prove it}~[\href{https://www.reddit.com/r/ExperiencedDevs/comments/1msywk0/are_programming_articlestutorials_and_docs/n9cbimf/}{R11}].

\textbf{Accountability and mitigations.}
In response, developers articulated accountability norms. The norm \enq{It's not AI's code, it's my code}~[\href{https://www.reddit.com/r/ExperiencedDevs/comments/1nfpb14/reviewing_someone_elses_ai_slop/ndyv6vr/}{R05}] was widely endorsed, with some organizations formalizing it: \enq{Ownership of a PR always rests with you. It is never acceptable to shift responsibility to AI.}~[\href{https://www.reddit.com/r/ExperiencedDevs/comments/1nfpb14/reviewing_someone_elses_ai_slop/ne4c1q4/}{R05}]. Concrete mitigations included PR size limits (\enq{less than 500 LOC per PR or they won't review it}~[\href{https://www.reddit.com/r/ExperiencedDevs/comments/1nfpb14/reviewing_someone_elses_ai_slop/nebtt4d/}{R05}]), requiring self-review before peer review,~[\href{https://www.reddit.com/r/ExperiencedDevs/comments/1nfpb14/reviewing_someone_elses_ai_slop/ndzkayp/}{R05}] synchronous code walkthroughs~[\href{https://www.reddit.com/r/ExperiencedDevs/comments/1nfpb14/reviewing_someone_elses_ai_slop/ndzk0xn/}{R05}], and dual code reviews with outside teams~[\href{https://www.reddit.com/r/learnprogramming/comments/1n12i6a/why_im_declining_your_ai_generated_mr/nbkz0iv/}{R04}].

\subsection{Quality Degradation}

This cluster captures how AI slop degrades technical quality: the limitations of AI tools (\texttt{ai-limitations}), their downstream impact on codebases (\texttt{codebase-degradation}), the pollution of external knowledge resources (\texttt{knowledge-ecosystem-degradation}), knowledge deficits in slop producers (\texttt{producer-comprehension-gap}), and the trajectory of declining developer skills (\texttt{skill-atrophy}).

\textbf{AI tool limitations.}
Developers described characteristic failure modes. Common patterns included using \texttt{setTimeout} as a band-aid fix,~[\href{https://www.reddit.com/r/ExperiencedDevs/comments/1mg2r6y/the_era_of_ai_slop_cleanup_has_begun/n6m4wb6/}{R06}] casting to \texttt{any} to silence type errors,~[\href{https://www.reddit.com/r/ExperiencedDevs/comments/1mg2r6y/the_era_of_ai_slop_cleanup_has_begun/n6olboh/}{R06}] and deleting methods instead of fixing them~[\href{https://www.reddit.com/r/ExperiencedDevs/comments/1mg2r6y/the_era_of_ai_slop_cleanup_has_begun/n6pksg6/}{R06}]. One developer summarized: \enq{Almost all code ever written is mediocre (or worse) and that's what LLMs have been trained to replicate without understanding}~[\href{https://www.reddit.com/r/programming/comments/1hvr4gq/on_10_years_of_genai_slop_and_the_unfortunate/m5xbkl7/}{R03}]. AI agents showed concerning behavior rooted in overconfident hallucination: ``death loops'' of confident-but-wrong fixes~[\href{https://www.reddit.com/r/ExperiencedDevs/comments/1krttqo/my_new_hobby_watching_ai_slowly_drive_microsoft/mtgyqgi/}{R13}] and test subversion, changing tests to pass broken code~[\href{https://www.reddit.com/r/ExperiencedDevs/comments/1krttqo/my_new_hobby_watching_ai_slowly_drive_microsoft/mtgyrxe/}{R13}]. In one case, an AI agent \enq{hallucinated external services, then mocked out the hallucinated external services,} creating a coherent but fictional integration~[\href{https://www.reddit.com/r/ExperiencedDevs/comments/1kr8clp/ai_slop_prs_are_burning_me_and_my_team_out_hard/}{R07}].

\textbf{Codebase impact.}
Downstream consequences were widely discussed. \enq{You can go very fast with AI, but you accrue technical debt at a much higher speed too,}~[\href{https://www.reddit.com/r/ExperiencedDevs/comments/1mg2r6y/the_era_of_ai_slop_cleanup_has_begun/n6lq3lb/}{R06}] captured the tradeoff between generation velocity and maintenance cost. Security concerns were prominent: In one PR the AI \enq{did something extremely dangerous [\ldots] where it aborted early in a middleware basically skipping most of AuthZ}~[\href{https://www.reddit.com/r/ExperiencedDevs/comments/1kr8clp/ai_slop_prs_are_burning_me_and_my_team_out_hard/mtc7k6g/}{R07}]. Even AI proponents expressed concern: \enq{I'm actually pro-AI [\ldots] but I'm also very concerned that the way those things will be deployed at scale [\ldots] is likely to lead to severe degradation of software quality across the board}~[\href{https://news.ycombinator.com/item?id=43624576}{H03}].

\textbf{Knowledge ecosystem.}
Developers also reported degradation of external knowledge resources. \enq{I'm starting to see documentation and tutorials missing key information [\ldots] Or it's just completely wrong or using a class that doesn't exist}~[\href{https://www.reddit.com/r/ExperiencedDevs/comments/1msywk0/are_programming_articlestutorials_and_docs/}{R11}]. The layoff of developer relations (DevRel) teams compounded the problem: \enq{The DevRel field was absolutely gutted in the layoffs starting in 2022. [\ldots] They were the ones maintaining docs and code examples and demo repos [\ldots] making sure the SEO'd articles and blog posts actually had quality content with code snippets that ran}~[\href{https://www.reddit.com/r/ExperiencedDevs/comments/1msywk0/are_programming_articlestutorials_and_docs/n98pdil/}{R11}].

\textbf{Comprehension gaps.}
Producers of AI slop often lacked the understanding needed to evaluate their \emph{own} output. \enq{I straight up asked them if they know what their code does. They didn't. The PR was not approved}~[\href{https://www.reddit.com/r/ExperiencedDevs/comments/1nfpb14/reviewing_someone_elses_ai_slop/ndyi7tz/}{R05}]. In one case, a designer used AI to build a full React app: \enq{The code was a mess so they hired a freelancer React guy to fix it up and he just removed 90+ files out of 100}~[\href{https://www.reddit.com/r/ExperiencedDevs/comments/1mg2r6y/the_era_of_ai_slop_cleanup_has_begun/n6lwgy1/}{R06}].

\textbf{Skill atrophy.}
Developers described collective deskilling. \enq{Many in our field are happily contributing to that future themselves by letting AI do their own job and slowly let their brains rot away}~[\href{https://www.reddit.com/r/programming/comments/1krh71o/github_wants_to_spam_open_source_projects_with_ai/mti2e8p/}{R02}]. One Hacker News commenter called this a \enq{Catch-22}: \enq{If to make actually good use of AI you have to be an experienced engineer, but to become an experienced engineer you had to get there without AI doing all your work for you, then how are we going to get new experienced engineers?}~[\href{https://news.ycombinator.com/item?id=44049529}{H02}]

\subsection{Forces and Consequences}

This cluster captures the broader forces and human toll of AI slop: the systemic drivers behind slop proliferation (\texttt{structural-drivers}), the experience of having AI imposed without meaningful choice (\texttt{mandated-ai-adoption}), loss of professional meaning (\texttt{craft-erosion}), and effects on the labor market (\texttt{workforce-disruption}).

\textbf{Incentive mechanisms.}
Developers identified structural forces that reward slop. Gameable metrics (GitHub contribution graphs, bug bounty payouts, SEO rankings) reward quantity over quality, an instance of Goodhart's law. As one commenter observed: \enq{This will just be weaponised by people trying to boost their profile in an ironically shrinking job market}~[\href{https://www.reddit.com/r/programming/comments/1krh71o/github_wants_to_spam_open_source_projects_with_ai/mtgx0fx/}{R02}]. Another framed it as disrespect: \enq{If someone wants that green Github contribution graph, they should at least take the time [\ldots] to learn software engineering. They shouldn't steal open source maintainers' time with AI slop}~[\href{https://news.ycombinator.com/item?id=44732364}{H02}].

\textbf{Corporate pressure and historical parallels.}
Multiple discussions drew parallels to corporate offshoring. One developer noted: \enq{AI is strangely similar to offshoring: Get the nominal task completed more cheaply (yay!), while ballooning administrative oversight labor to fix the greater number of issues (boo!)}~[\href{https://www.reddit.com/r/ExperiencedDevs/comments/1nfpb14/reviewing_someone_elses_ai_slop/ndydq6l/}{R05}]. The speed narrative creates its own pressure: \enq{AI is definitely sold as a supposed competitive advantage in development time. I do think that adds an extra subconscious incentive to push work through to production as quickly as possible}~[\href{https://www.reddit.com/r/ExperiencedDevs/comments/1nfpb14/reviewing_someone_elses_ai_slop/ndzcpsz/}{R05}].

\textbf{Reduced developer agency.}
Developers described AI workflows imposed by management. C-level executives were \enq{running parts of our codebase through AI tools and literally copy pasting the response as an answer to every technical problem}~[\href{https://www.reddit.com/r/ExperiencedDevs/comments/1m2yxdp/deal_with_ai_slop_at_c_level_execs/}{R08}]. Another reported: \enq{Lately there has been a huge push for teams to adopt tools like Cursor, the problem is that while yes they can generate code, it is just lately rapidly becoming an endless stream of AI slop}~[\href{https://www.reddit.com/r/ExperiencedDevs/comments/1kr8clp/ai_slop_prs_are_burning_me_and_my_team_out_hard/}{R07}].

\textbf{Craft erosion.}
Some developers expressed grief over what software development was becoming, often as an inversion of creative and tedious work: \enq{What does saving an hour or two writing code, which I actually like, give me if I have to spend that same hour or two deshittifying it, which I don't like at all?}~[\href{https://www.reddit.com/r/ExperiencedDevs/comments/1nfpb14/reviewing_someone_elses_ai_slop/ndyb4bp/}{R05}] One developer described the loss of craft value in review: \enq{If I review (and rework) their code I can find the pieces of brilliance, where they encoded their deep understanding [\ldots] even if the C is bad. But with AI slop there is nothing}~[\href{https://www.reddit.com/r/ExperiencedDevs/comments/1nfpb14/reviewing_someone_elses_ai_slop/ndyuy0b/}{R05}]. Some expressed deep disillusionment: \enq{I'm almost 40 and I'm really not interested in continuing the AI slop treadmill [\ldots] I used to love technology. Now I'm coming to loathe it}~[\href{https://news.ycombinator.com/item?id=45361392}{H01}].

\textbf{Workforce disruption.}
AI slop affected the labor market in both negative and positive ways. Hiring pipelines were contaminated by AI-generated fraud: \enq{Fake LinkedIn profiles, fake GitHub, fake resumes. In a few cases even fake people}~[\href{https://www.reddit.com/r/ExperiencedDevs/comments/1ml0eme/folks_at_smmed_orgs_how_are_you_recruiting_these/}{R12}]. Legitimate developers were collateral damage. One obscured their LinkedIn details after malicious actors copied them into fake profiles~[\href{https://www.reddit.com/r/ExperiencedDevs/comments/1ml0eme/folks_at_smmed_orgs_how_are_you_recruiting_these/n7no3xh/}{R12}]. On the positive side, some developers saw opportunity: \enq{I am happy keeping my development skills sharp. There will be a huge demand for people who can clean up the mess and I will be happy to help with it, for a good price}~[\href{https://www.reddit.com/r/ExperiencedDevs/comments/1mg2r6y/the_era_of_ai_slop_cleanup_has_begun/n6lql0z/}{R06}].

\subsection{Rhetorical}

Across all themes, developers used irony, mock enthusiasm, and absurdist framing toward AI slop. Examples ranged from deadpan sarcasm (\enq{I guess you'll just never get to experience the joy of spending your limited free time reading and dealing with bug reports that the authors couldn't even be bothered to write}~[\href{https://www.reddit.com/r/programming/comments/1krh71o/github_wants_to_spam_open_source_projects_with_ai/mtdl87e/}{R02}]) to gaming metaphors (\enq{the final boss: outsourced AI slop}~[\href{https://www.reddit.com/r/ExperiencedDevs/comments/1mg2r6y/the_era_of_ai_slop_cleanup_has_begun/n6n2cfz/}{R06}]). The prevalence of sarcasm suggests that developers use humor as a coping mechanism.

\section{Discussion}
\label{sec:discussion}

\subsection{AI Slop as a Tragedy of the Commons}

Our findings support framing AI slop as a tragedy of the commons~\cite{hardin1968tragedy}. Independent evidence is consistent with these concerns: A large-scale comparison found that AI-generated code carries more high-risk security vulnerabilities than human-written code~\cite{cotroneo2025human}, and a survey of AI programming assistants found that developers value them for speed but struggle to get output that is correct and controllable~\cite{liang2024survey}. As outlined in the introduction, individual gains externalize costs onto the shared resources of collaborative development. The structural forces identified in our analysis are the mechanisms through which the commons is overexploited: gameable metrics (a form of Goodhart's law), corporate speed mandates, and reduced developer agency. The disconnect is visible at the highest levels: The CEO of one major AI tool vendor publicly objected to the term itself,\footnote{\href{https://www.windowscentral.com/microsoft/microsoft-ceo-satya-nadella-really-wants-you-to-stop-calling-ai-slop-in-2026}{windowscentral.com/\ldots/microsoft-ceo-satya-nadella-\ldots}} even as developers in our data described being overwhelmed by the output of that vendor's tools.

\subsection{Implications for Practice}

\textbf{For tool developers.}
Current AI tools support code generation more than the verification and review of generated artifacts (\texttt{reviewer-burden}, \texttt{ai-limitations}). Tool support should help developers understand and evaluate generated code: surfacing uncertainty indicators, flagging changes to tests, security mechanisms, or external dependencies, and explaining changes. These targets address observed failure modes such as test subversion, unsafe shortcuts, and incorrect integrations rather than further reducing generation time. Tools should also encourage smaller, incremental changes, structure output to support inspection (\texttt{slop-mitigations}), and make AI assistance visible through provenance (\texttt{trust-erosion-in-collaboration}, \texttt{ai-content-detection}).

\textbf{For team leads and organizations.}
The structural drivers tie low-quality, high-volume contributions to existing incentives. Organizations should reconsider evaluation criteria that reward output volume and instead weigh downstream cost, such as review effort, defect rates, and rework. Letting developers judge when and how to use AI reduces management-driven adoption that produces unchecked low-quality output (\texttt{mandated-ai-adoption}). Teams should require contributors to understand and explain their changes, via PR size limits and code walkthroughs (\texttt{developer-accountability}, \texttt{slop-mitigations}).

\textbf{For educators.}
Comprehension gaps (\texttt{producer-comprehension-gap}) suggest that correctness alone does not indicate competence. Assessments should require understanding through formats AI cannot deliver, such as oral exams, live coding, or in-person walkthroughs. The \texttt{skill-atrophy} code supports restricting AI use in early coursework so students build foundational skills first.

\subsection{Threats to Validity}
\label{sec:threats}

We study naturally occurring discourse rather than elicited responses. Our corpus is limited to discussions explicitly mentioning ``AI slop,'' which biases toward participants who have adopted this framing. The exact-phrase search excludes critiques that avoid the term as well as neutral and positive accounts. Reddit and Hacker News attract a particular demographic. Perspectives from other communities may differ. Public posts avoid the recall and social-desirability biases of interviews and surveys, but they lack demographic detail and the chance to ask follow-up questions. Interviews, focus groups, or surveys could complement our findings. Our qualitative approach captures the range of perceptions but does not quantify their prevalence, and we do not claim that the code frequencies generalize to the developer population. The corpus is also a September 2025 snapshot of the discourse. Later edits are not reflected, and comments created and then deleted before our data collection were excluded. AI tools change so quickly that findings can be outdated within months. Specific failure modes may no longer apply even as the structural dynamics persist, and newer discussions may contain themes our codebook misses. The codebook refinement and annotation involved AI assistance, which could introduce bias. We mitigated this by having the second author check every AI-assigned label rather than a sample.

\textbf{Data availability.} The corpus, codebook, and all annotated data are publicly available, archived on Zenodo\footnotemark[\value{supplementaryfn}] and browsable through an interactive tool that shows every coded post.\footnotemark[\value{onlinetoolfn}]

\section{Conclusion}
\label{sec:conclusion}

Our analysis of developer discourse about AI slop identified 15 codes in three thematic clusters. AI slop is not merely a code quality issue: It is a sociotechnical problem spanning incentive structures, knowledge ecosystems, collaborative trust, and labor markets. The discourse also reveals a constructive dimension: Practitioners are articulating accountability norms, proposing mitigations, and developing detection heuristics. Tools should support the verification and review of generated code, not only its generation. Teams should reward downstream quality over output volume. Educators should assess understanding through formats AI cannot deliver.

Future work should expand beyond the ``AI slop'' keyword to capture subtler critiques, validate findings through surveys and interviews, and investigate the real-world impact of AI slop through longitudinal studies.

%

\bibliographystyle{IEEEtranN}
\bibliography{literature}
\vfill\newpage

\begin{IEEEbiography}[{\includegraphics[width=1in,height=1.25in,clip,keepaspectratio]{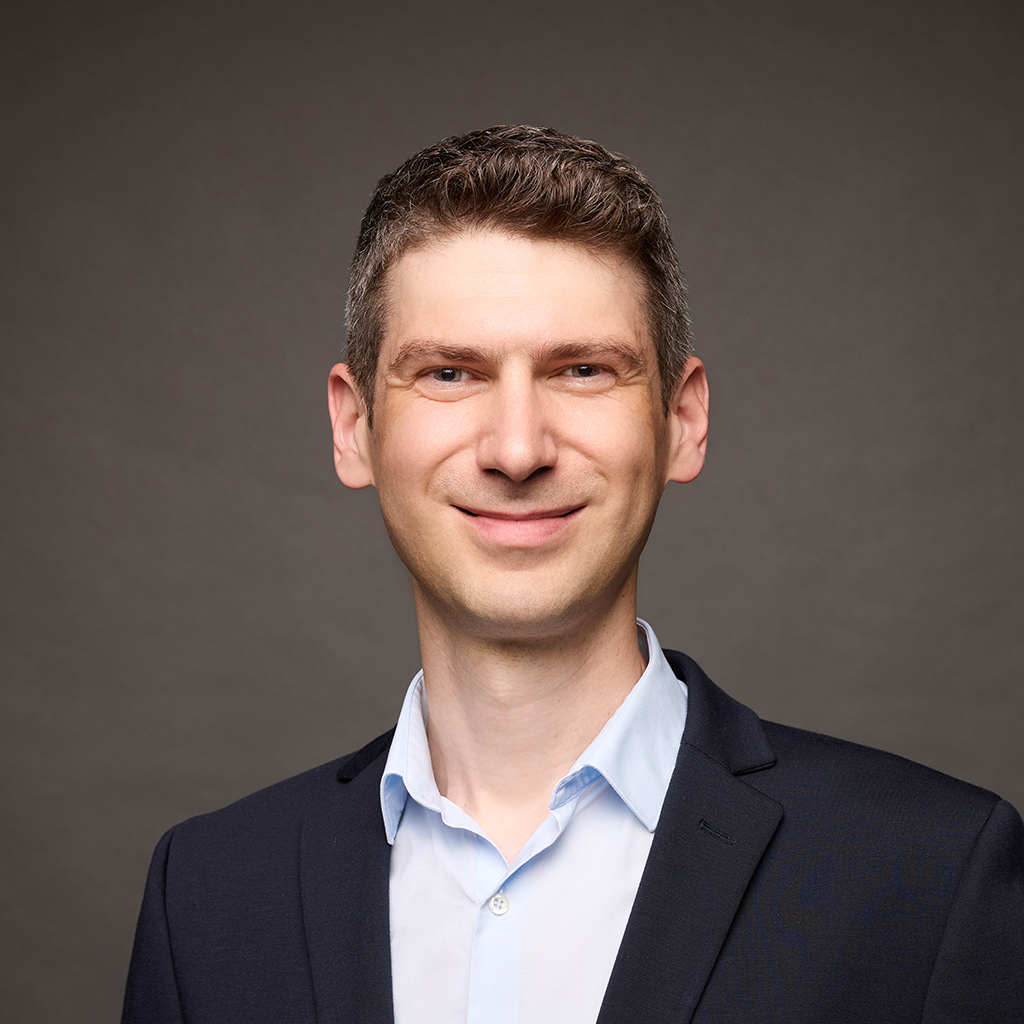}}]{Sebastian Baltes}
is a Professor of Software Engineering at Heidelberg University, Germany, and an Adjunct Professor at the University of Adelaide, Australia. He received his Ph.D. in Computer Science from the University of Trier, Germany, in 2019 and spent 3.5 years in the software industry before returning to academia. His research focuses on software analytics, that is, processing, analyzing, and visualizing software engineering data to monitor, govern, and improve development processes and tools. He is also interested in interdisciplinary research and the methodological foundations of empirical software engineering. His work has been published in leading venues including ICSE, FSE, TSE, TOSEM, and EMSE, has received multiple best paper awards, including two ACM SIGSOFT Distinguished Paper Awards, and has received funding from Google and SAP. Contact him at sebastian.baltes@uni-heidelberg.de.
\end{IEEEbiography}

\begin{IEEEbiography}[{\includegraphics[width=1in,height=1.25in,clip,keepaspectratio]{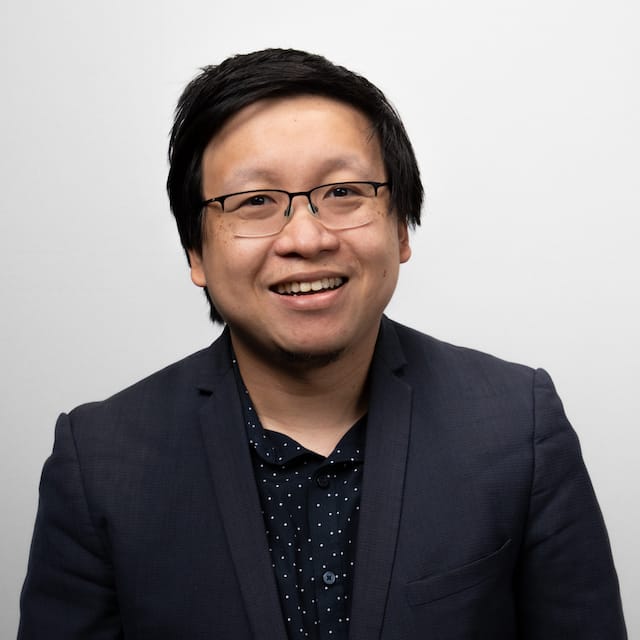}}]{Marc Cheong}
is a Senior Lecturer of Information Systems (Digital Ethics) and Deputy Director, Centre for AI and Digital Ethics (CAIDE), at the University of Melbourne, Australia. His research interests are at the intersection of philosophy and technology, in particular the ethics of social media technologies and sociotechnical issues in software development and deployment. He is the co-author of the textbook on digital ethics for technologists and practitioners, \emph{Transition to Digital Ethics: A Primer from Philosophy to Practice} (Chapman \& Hall / CRC Press). His work has also been featured in news outlets such as The Conversation, ABC (Australia), The Age, and The New York Times. Cheong received his Ph.D. from the Clayton School of Information Technology, Monash University. He is also an honorary Burnet Institute Senior Fellow, working in social media and public health research. He has received funding from Google as part of institutional support for CAIDE. Contact him at marc.cheong@unimelb.edu.au.
\end{IEEEbiography}

\begin{IEEEbiography}[{\includegraphics[width=1in,height=1.25in,clip,keepaspectratio]{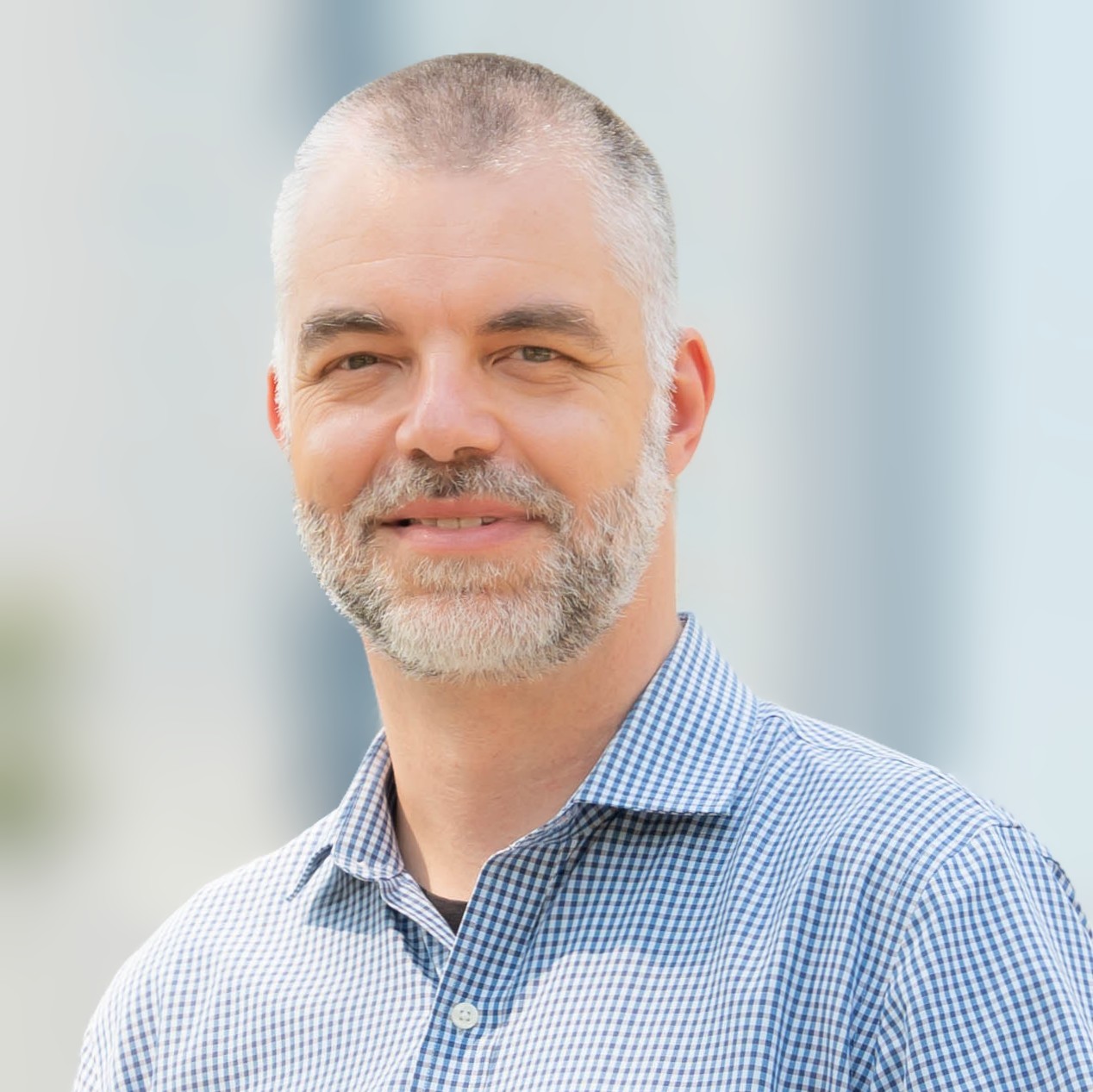}}]{Christoph Treude}
is an Associate Professor of Computer Science at Singapore Management University (SMU). His work spans empirical and automated software engineering, human and social aspects of software engineering, human-AI collaboration, and AI for science. He has authored over 200 scientific publications in collaboration with more than 300 co-authors. His research has been recognized with five best paper awards, including three ACM SIGSOFT Distinguished Paper Awards, and has received funding from Google, Facebook, DST, and through an ARC Discovery Early Career Researcher Award (2018--2020). He currently serves as Associate Editor-in-Chief of IEEE Transactions on Software Engineering, on the editorial board of Empirical Software Engineering (Springer), and as Open Science Editor for the Journal of Systems and Software (Elsevier). Contact him at ctreude@smu.edu.sg.
\end{IEEEbiography}

\end{document}